\documentclass[nofootinbib,showpacs,pra,twocolumn,aps,floatfix]{revtex4}
\usepackage{bm}
\usepackage{amssymb}
\usepackage{graphicx}
\usepackage{mathtext}
\usepackage{amsfonts}
\usepackage{mathrsfs}
\def\d{{\rm d}}
\def\la{\langle}
\def\ra{\rangle}
\def\om{\omega}

\newcommand{\beq}{\begin{equation}}
\newcommand{\eeq}{\end{equation}}
\newcommand{\beqa}{\begin{eqnarray}}
\newcommand{\eeqa}{\end{eqnarray}}

\newcommand{\ket}[1]{|#1\rangle}             
\newcommand{\bra}[1]{\langle #1|}            
\renewcommand{\d}[1]{\mathrm{d}#1}           
\newcommand{\mt}[1]{\mathrm{#1}}             
\newcommand{\bgl}[1]{\boldsymbol{#1}}         

\begin{document}

\title{Ramsey interferometry with a two-level Tonks-Girardeau gas}


\author{S. V. Mousavi}
\email{s_v_moosavi@mehr.sharif.edu}
\affiliation{Departamento de Qu\'\i mica-F\'\i sica, 
Universidad del Pa\'\i s Vasco, Apdo. 644,
48080 Bilbao, Spain}
\affiliation{Department of Physics, Sharif University of Technology, P. O. Box 11365-9161, Tehran, Iran}
\author{A. del Campo}
\email{adolfo.delcampo@ehu.es}
\affiliation{Departamento de Qu\'\i mica-F\'\i sica, Universidad del Pa\'\i s Vasco, Apdo. 644,
48080 Bilbao, Spain}
\author{I. Lizuain}
\email{ion.lizuain@ehu.es}
\affiliation{Departamento de Qu\'\i mica-F\'\i sica, Universidad del Pa\'\i s Vasco, Apdo. 644,
48080 Bilbao, Spain} 
\author{J. G. Muga}
\email{jg.muga@ehu.es}
\affiliation{Departamento de Qu\'\i mica-F\'\i sica, Universidad del Pa\'\i s Vasco, Apdo. 644,
48080 Bilbao, Spain} 

%
\begin{abstract}
We propose a generalization of the Tonks-Girardeau model that describes 
a coherent gas of cold two-level Bosons which interact with two external fields
in a Ramsey interferometer. They also interact among themselves by contact collisions with interchange of momentum and internal state.
We study  
the corresponding Ramsey fringes and the quantum projection noise which, essentially unaffected by the interactions, remains that for ideal Bosons. The dual system of this gas, an ideal gas of two-level Fermions coupled by the interaction with the separated fields, produces the same fringes and noise fluctuations. The cases of time-separated and spatially-separated fields are studied. For spatially separated fields the  
fringes may be broadened slightly by increasing the number of particles, but only for large particle numbers far from present experiments with Tonks-Girardeau gases. The 
uncertainty in the determination of the atomic transition frequency diminishes, 
essentially with the inverse root of the particle number. 
\end{abstract}
\maketitle
\section{Introduction}
%
 
A basic feature of the observed interference fringes in a standard Ramsey experiment
is that their width is determined by the inverse of the time taken by the atoms to cross the intermediate drift region. For precision measurement purposes, as in atomic clocks, this motivates the use of very slow (ultracold) atoms and therefore the development of laser cooling techniques has changed the entire prospects of frequency standards \cite{22}. Experimentally, atomic velocities of the order of $1\,\mt{cm/s}$ and smaller can be achieved, and space-based clocks are in development to eliminate gravitational effects in the motion of such slow particles \cite{Salomon}.
Laser cooled atoms are also interesting in metrology and interferometry because of the possibility to achieve narrow velocity distributions and avoid averaging effects. In addition, fundamentally new effects may arise by using coherent few-body or many-body 
states as input in the form of condensates or otherwise: for example, there exist proposals to beat the limitations imposed by quantum projection noise using entanglement \cite{Wineland,Cirac}.   

In spite of the above, the motto ``the slower the better'' in the context of atomic 
clocks has actually a limited domain beyond which quantum motional phenomena may affect strongly and eventually deform totally the usual Ramsey pattern. 
If the slow atom moves initially along the $x$-axis and the fields are oriented perpendicularly along the $y$-axis, there are two origins of modification of the standard Ramsey result \cite{Ramsey}. First, the absorption of a photon leads to a transverse momentum transfer on the atom, such that the excited state separates in space from the ground state.
This is negligible for microwaves but not for optical transitions.  The effect can be understood classically by means of energy conservation and momentum conservation in $y$-direction. It has been studied in detail by Bord\'e and coworkers  \cite{Borde} and multi-beam setups have been implemented to correct for this separation in order to observe quantum interference \cite{Bo-PRA-1984, KaChu, Mo-PRA-1992}. 
Second, the field acts as a barrier for the longitudinal motion of the atom, and quantum reflection and tunneling may occur. Thus, momentum in $x$-direction is not conserved as a consequence of the $x$-dependence of the fields. 
For microwave fields and the corresponding Rabi frequencies these quantized motion effects are tiny for present atomic velocities but may become important for deeply ultracold  particles. Moreover, in view of a
proclaimed near-future accuracy of frequency standards of $10^{-18}$\cite{Schiller-preprint}, even those tiny effects have to be studied beyond the limits of the standard theoretical description of the Ramsey pattern.
Reflection effects have been considered for ultracold atoms passing through one \cite{Mazer1} or two \cite{Mazer2} resonant micromaser cavities, leading to the 
concept of ``mazer physics". The nonresonant single-mazer case has been investigated by 
Bastin and Martin \cite{BaMa-PRA-2003}. Recently, we extended the study to two separated classical fields and 
gave an exact quantum result of the Ramsey fringes as a function of the detuning including quantum tunneling and reflection by means of two-channel recurrence relations \cite{SeiMu-EPJD}. 

Apart from quantum motion effects affecting ensembles of independent particles,
other effects are due to the importance of quantum statistics and interactions.
The use of a Bose-Einstein condensate for an atomic clock immediately comes to mind, but
the improvements associated with low velocities and narrow velocity distribution may be 
compensated by negative effects, such as collisional shifts and 
instabilities leading to the separation of the gas cloud \cite{Cornell02,Band06}.

A natural candidate for further exploration is the Tonks-Girardeau (TG) regime
of impenetrable, tightly confined Bosons subjected to hard-core ``contact'' interactions  \cite{Gir60,Gir65}, since some of its properties are in a sense opposite to those of the condensate.   
In particular, the TG requirement of contact interactions, implies strong similarities 
between the Bosonic system and a ``dual'' system of freely moving Fermions, 
with all local correlation functions of both being actually equal. 
Other important feature of the TG gas is its one dimensional (1D) character.   
Olshanii showed \cite{Ols98,BerMooOls03,PetShlWal00} that when 
a Bosonic vapor
is confined in a de Broglie wave guide with transverse trapping so tight and 
temperature so low that the transverse vibrational excitation quantum 
$\hbar\omega_{\perp}$ is larger than available longitudinal zero point and 
thermal energies, the effective dynamics becomes one dimensional, 
and accurately described by a 1D Hamiltonian with delta-function interactions
$g_{1D}\delta(x_j-x_{\ell})$, where $x_j$ and $x_{\ell}$ are 1D longitudinal
position variables. This is the Lieb-Liniger (LL) 
model, exactly solved in 1963 by a Bethe ansatz method \cite{LieLin63}.
The coupling constant $g_{1D}$ can be tuned using a Feschbach resonance, allowing to reach 
the Tonks-Girardeau regime of impenetrable Bosons, which corresponds to the  
$g_{1D}\to\infty$ limit of the LL model.
It has been  realized experimentally \cite{Par04,Kin04}, and was solved
exactly in 1960 \cite{Gir60,Gir65} by the so called Fermi-Bose mapping to
the ideal Fermi gas.   

For metrology and interferometry applications the tight 1D confinement
along a waveguide
is a simplifying feature since no transversal motional branches have to 
be considered with the possible bonus of an increased signal.
(The spatial separation into several branches may also be desirable, as in Sagnac
interferometry, but it could be implemented with waveguides too.)  
Nevertheless, the confinement is by itself problematic for frequency standard 
applications, since it is carried out by means 
of magnetic or optical interactions which will in principle perturb the internal
state levels of the atom. Several schemes have been proposed
to mitigate this problem and compensate the shifts due  
to magnetic \cite{Cornell02} or optical interactions \cite{ss,Ha}.\footnote{We shall assume hereafter that such compensation is implemented.}       
       
The possible applications in interferometry are a strong motivation for current research 
in TG gases. Interference effects have been examined so far in a few publications   
in which internal states have not played any role \cite{Gir00,Gir02}.
Indeed, a TG model
including internal states and an external interaction coupling them has not
been discussed, although 
optically guided systems with free spin subjected to potentials for singlet and triplet 
interactions have been studied by means of effective LL models
\cite{GirNguOls04,Gir06b}. Note also that a model applicable to a two-level LL gas coupled 
by an on-resonance laser has been solved by nested Bethe ansatz \cite{Yang67}.
  
In this paper, we investigate the implications in Ramsey interferometry of a model in the spirit of the original (structureless) TG 
gas but with internal structure. The interactions 
defined allow us to achieve essential solvability of the dynamical problem in the Ramsey two-field excitation setup
by simple quadrature:
the collisions are characterized by internal state and momentum exchange, which reduce to the usual impenetrable constraint for collisions in the same internal channel.

We shall consider different configurations for the two fields, both in space and time domains. 
They are conceptually different and the mathematical treatment is different 
too. For reasonable parameters, however, the 
results turn out to be very similar.    
\section{Two-level Tonks-Girardeau gas with exchange, contact interactions}
We shall propose here a generalization of the Tonks-Girardeau gas for two-level impenetrable atoms. First we shall need to review or introduce some notation and basic concepts.  
In one dimension the state of a single two-level atom may be written as 
\beqa
\Phi_n(x_1)=
\sum_{b=g,e}\phi^{(b)}_n(x_1)\vert b\ra, 
\eeqa
%
where $n=1,2,3...$ is a label to distinguish different wave functions 
and $b$ is a generic index for the internal bound state which may be $g$ (ground), 
or $e$ (excited).
One-particle states may be combined to form two-particle ones with the form 
\beq
\Phi_{nn'}(x_1,x_2)=
\sum_{b,b'}\phi_n^{(b)}(x_1)\phi_{n'}^{(b')}(x_2)|bb'\ra, 
\eeq
and similarly for more particles. 
The convention in  $|bb'\ra$ is that $b$ is for particle 1 and 
$b'$ for particle 2. This will in some equations be indicated even more explicitly 
adding a particle subscript to the internal state label, $b_1$, $b_2$, etc.

Consider now the usual Pauli operators acting on one-particle internal state vectors,
\beqa
\sigma_{X}&=&|g\ra\la e|+|e\ra\la g|,
\nonumber\\
\sigma_Y&=&i(|g\ra\la e|-|e\ra\la g|),
\nonumber\\
\sigma_Z&=&|e\ra\la e| -|g\ra\la g|,
\eeqa
and the corresponding 3-component operator 
$\hat{\mathbf{S}}_j=\bgl{\sigma}_j/2$ for particle $j$
analogous to the spin-$1/2$ angular momentum operator.  
If $\mathbf{S}=\mathbf{S}_1+\mathbf{S}_2$,
$\mathbf{S}^2$ has eigenvalues $S(S+1)$ with $S=0$ and $S=1$
corresponding to singlet and triplet subspaces as it is 
well known.  

Assume now the following Hamiltonian
\begin{equation}\label{Fermi Hamiltonian_2}
\hat{H}_{\rm{coll}}=-\frac{\hbar^2}{2m}\sum_{j=1}^{2}\partial_{x_j}^{2}
+v_s(x_{12})\hat{P}_{12}^s
+v_{t}(x_{12})\hat{P}_{12}^t.
\end{equation}
Here $x_{12}=x_1-x_{2}$, 
and $\hat{P}_{12}^s=\frac{1}{4}-\hat{\mathbf{S}}_1\cdot\hat{\mathbf{S}}_{2}$ 
and 
$\hat{P}_{12}^t=\frac{3}{4}+\hat{\mathbf{S}}_1\cdot\hat{\mathbf{S}}_{2}$
are the projectors onto the subspaces of singlet and triplet functions.  
      
The internal Hilbert space can be written as $\mathcal{H}_s\oplus\mathcal{H}_t$, 
where $\mathcal{H}_s$ is spanned by $(|eg\ra-|ge\ra)/\sqrt{2}$ and $\mathcal{H}_t$ by $\{|gg\ra,|ee\ra,(|eg\ra+|ge\ra)/\sqrt{2}\}$. 
Suppose that the reflection amplitude for relative motion 
in such representation takes the values $+1,-1$ 
in singlet and triplet subspaces respectively. The particles are impenetrable 
and these values correspond to a hard wall potential $v_t$,  
whereas $v_s$ is a hard-core repulsive potential with an additional well of width $l$ and 
depth $V$,  
so that the reflection amplitude becomes $R=+1$ in the limit in which the well is made infinitely narrow and the well 
infinitely deep, keeping $(2mV/\hbar^2)^{1/2} l=\pi/2$ \cite{GirOls03,GirOls04,GirNguOls04,CheShi98}. 
Translated into the bare basis this implies that in all collisions between atoms in $g$ or $e$ and
well defined momenta, they interchange their momenta 
(the relative momentum changes sign), as well as their internal state,
with the outgoing wave
function picking up a minus sign because of the hard-core reflection. 
For $x_1<x_2$ and equal internal states such collision is represented by
\beq
e^{ikx_1}e^{ik'x_2}|bb\ra-[e^{ik'x_2}e^{ikx_1}]|bb\ra,
\label{diag}
\eeq
whereas for $b\ne b'$,
\beq
e^{ikx_1}e^{ik'x_2}|bb'\ra-[e^{ik'x_2}e^{ikx_1}]|b'b\ra.
\label{ndiag}
\eeq
In the diagonal case of equal internal states the spatial part vanishes at contact, $x_1=x_2$,
whereas in the 
non-diagonal case it does not, but note that in Eq. (\ref{ndiag}) only the 
``external region'' is considered, disregarding the infinitely narrow well region.    
%
\begin{figure}
\includegraphics[width=5cm,angle=0]{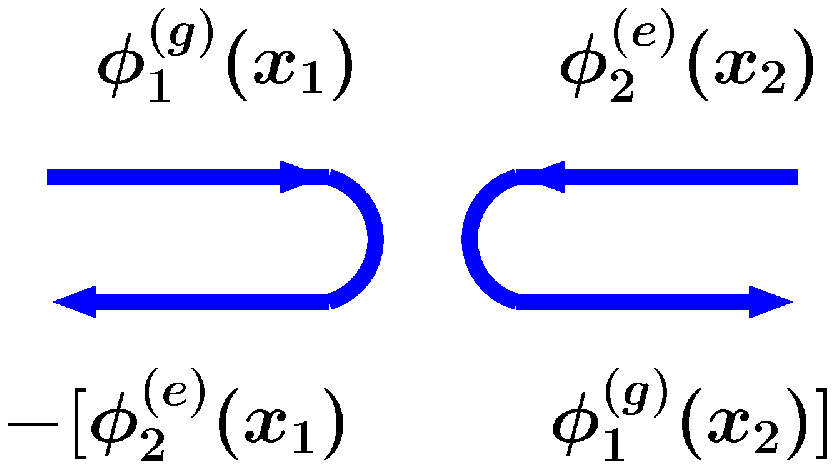} 
\includegraphics[width=5cm,angle=0]{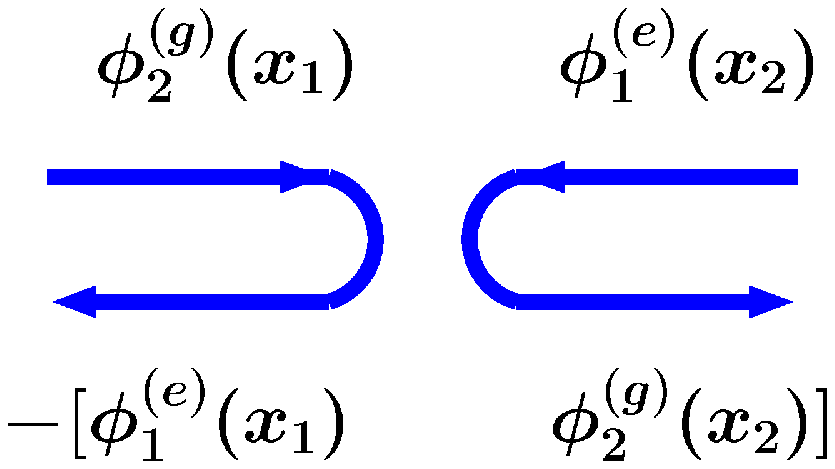} 
  \caption{Diagrammatic representation of the collisions for particles on 
different channels.  Note that particles do not cross and interchange their 
momentum and internal state picking up an additional phase (minus sign).
} \label{coll}
\end{figure}
%

Up to now we have made no reference to quantum statistics 
and the particles are formally distinguishable. 
Let us consider now a Fermionic state made of freely moving 
one-particle states with the form 
\beqa
\Psi_F(x_1,x_2)&=&\frac{1}{\sqrt{2}}{\rm det}_{n,m=1}^{2}\Phi_{n}(x_{m})
\nonumber\\
&=&\frac{1}{\sqrt{2}}\sum_{b_1,b_2=g,e}
\left\vert\begin{array}{cc}
\phi_1^{(b_1)}(x_1) & \phi_1^{(b_2)}(x_2) \\
\phi_2^{(b_1)}(x_1) & \phi_2^{(b_2)}(x_2)
\end{array}\right\vert
\vert b_1 b_2\ra,
\nonumber
\eeqa
with the state sign changing by switching particles 1 and 2 and the internal states.
 

%
%
More explicitly, the different terms can be rearranged as
\beqa
&&2^{1/2}\Psi_F(x_{1},x_{2})=
\nonumber\\
&+&[\phi_1^{(g)}(x_1)\phi_2^{(g)}(x_2)-\phi_2^{(g)}(x_1)\phi_1^{(g)}(x_1)]|gg\ra
\nonumber\\
&+&[\phi_1^{(e)}(x_1)\phi_2^{(e)}(x_2)-\phi_2^{(e)}(x_1)\phi_1^{(e)}(x_2)]|ee\ra
\nonumber\\
&+&[\phi_1^{(g)}(x_1)\phi_2^{(e)}(x_2)]|ge\ra -[\phi_2^{(e)}(x_1)\phi_1^{(g)}(x_2)]|eg\ra
\nonumber\\
&+&[\phi_1^{(e)}(x_1)\phi_2^{(g)}(x_2)]|eg\ra
-[\phi_2^{(g)}(x_1)\phi_1^{(e)}(x_2)]|ge\ra.
\eeqa
This form makes evident that the Fermionic character of the state  
imposes for diagonal or non-diagonal 
terms the contact boundary conditions specified above, compare with  
Eqs. (\ref{diag},\ref{ndiag}).
An associated Bosonic system, totally symmetric under $(x_i,b_i)\leftrightarrow(x_j,b_j)$ permutations 
may be now obtained 
by means of the Bose-Fermi mapping, 
$\Psi(x_1,x_2)=\mathcal{A}\Psi_F(x_1,x_2)$, where the 
antisymmetric unit function is 
$\mathcal{A}={\rm sgn}(x_{1}-x_{2})$.

We have in summary constructed a Bosonic wave function for a system of two particles subjected to 
contact interactions with internal state and momentum interchange,
using a dual system of two 
non-interacting Fermions and the antisymmetric unit function. 
The generalization to $N$-atoms is straightforward: 
\beqa
&&\Psi_F(x_1,\dots,x_N)=\frac{1}{\sqrt{N!}}{\rm det}_{n,m=1}^{N}\Phi_{n}(x_{m})
\nonumber\\
\!\!&\!\!=\!\!&\!\!
\frac{1}{\sqrt{N!}}\sum_{b_1, \cdots,b_N=g,e}
\left\vert\!\!\begin{array}{ccc}
\phi_1^{(b_1)}(x_1) & \cdots & \phi_1^{(b_N)}(x_N) \nonumber\\
\vdots & \ddots & \vdots \\
\phi_N^{(b_1)}(x_1) & \cdots & \phi_N^{(b_N)}(x_N)
\end{array}\!\!\right\vert\!
\vert b_1\cdots b_N\ra, 
\nonumber\eeqa
and 
\beq
\Psi(x_1,\dots,x_N)=\mathcal{A}\Psi_F(x_1,\dots,x_N),
\label{bos}
\eeq
where 
\beq 
\mathcal{A}=\prod_{1\leq j<k\leq N}{\rm sgn}(x_{k}-x_{j}), 
\eeq
is the Bosonic solution of the time-dependent or stationary Schr\"odinger equation 
for the Hamiltonian  
\begin{equation}\label{Fermi Hamiltonian}
\hat{H}_{\rm{coll}}=-\frac{\hbar^2}{2m}\sum_{j=1}^{N}\partial_{x_j}^{2}
+\sum_{1\le j<\ell\le N}[v_s(x_{j\ell})\hat{P}_{j\ell}^s
+v_{t}(x_{j\ell})\hat{P}_{j\ell}^t],
\end{equation}
with the same contact interactions as before.  

The density profile, normalized to $N$-particles,
which gives the appearance of the cloud, is defined by 
\beq
\rho_{N}(x)=N\int \parallel\Psi(x_{1},\cdots,x_{N})\parallel^2
\d x_2\cdots\d x_N.
\eeq
Provided that the one-particle states $\Phi_n$ are orthonormal,
as they will always be hereafter, 
the density profile reads 
\beq
\rho_N(x)  = 
\sum_{b=g,e}\sum_{n=1}^{N}\vert\phi_n^{(b)}(x)\vert^2
=\sum_{b=g,e}\rho_N^{(b)}(x),
\eeq
where the density profile for each of the channels defined by the two internal levels
is given by
\beq
\rho_N^{(b)}(x) = \sum_{n=1}^{N}\vert\phi_n^{(b)}(x)\vert^2.
\label{sumn}
\eeq
%

The simplicity achieved by our model parallels that of the usual (structureless) TG gas
in the sense that 
an $N-$body wavefunction with interactions is obtained from freely-moving one-body 
states. Even more, this property is preserved by adding an interaction affecting the individual
atoms only and coupling the internal 
levels. This is precisely the type of interaction that 
we find in the Ramsey interferometer.   

\section{Quantum projection noise in a two-level Tonks-Girardeau gas}
Itano and coworkers \cite{Itano93} studied the quantum projection noise 
for a Hartree product state of the form
$|b_1,\dots,b_N\ra=\otimes_{i=1}^{N}|b_i\ra$. This noise is, in other words, 
the fluctuation of the number of excited atoms for measurements made
in the $N$-body system. In Ramsey interferometry, the error in the determination of the atomic frequency 
depends on the ratio between the (root of the) fluctuation in the number of excited atoms and the 
derivative of the signal
(proportional to the number of excited atoms) with respect to detuning. 

Here we shall obtain the noise associated with the state $\Psi(x_{1},\cdots,x_{N})$. 
We shall follow \cite{Itano93} and introduce the operator
\beqa
\hat{S}_Z=\sum_{i=1}^N \hat{S}_{i_Z}=\frac{1}{2}\sum_{i=1}^N (|e_i\ra\la e_i|-|g_i\ra\la g_i|), 
\eeqa
where it is assumed, as usual, that each term in the summation is multiplied by the identity operator 
for all the other atoms. 

The quantum projection noise of a signal is 
proportional to the variance
\beqa
(\Delta S_Z)^2=\la \hat{S}_{Z}^{2}\ra-\la \hat{S}_Z\ra^2,
\eeqa
and expressions for both terms will now be worked out.   
First, notice that $S_Z$ commutes with ${\cal{A}}$ so that, using ${\cal{A}}^2$, we 
may compute the expectation values substituting $\Psi$ by 
$\Psi_F$, i.e., for the more easily tractable, dual Fermionic system.
Since $\Psi_F$ is antisymmetric it follows that
\beq
\la\hat{S}_Z\ra=N\la\Psi|\hat{S}_{1_Z}|\Psi\ra,
\label{sz}
\eeq
and
\begin{eqnarray}
\la \hat{S}_{Z}^{2}\ra&=&\sum_{i=1}^{N}\la\hat{S}_{i_Z}^{2}\ra+
\sum_{ i,j \,\, i\neq j}
\la\hat{S}_{i_Z}\hat{S}_{j_Z}\ra
\nonumber\\
&=&N\la\hat{S}_{1_Z}^{2}\ra
+N(N-1)\la\hat{S}_{1_Z}\hat{S}_{2_Z}\ra.
\label{sec}
\end{eqnarray}
Eq. (\ref{sz}), takes the form
\beqa
N\la\Psi|\hat{S}_{1_Z}|\Psi\ra\!\!&\!\!=\!\!&\!\!N\!\!\int\!\!\prod_i\! {\rm d}x_i\Psi^*(x_{1},\dots,x_{N})
\hat{S}_{1_Z}\Psi(x_{1},\dots,x_{N})
\nonumber\\&=&\sum_{n=1}^N\la\Phi_n|\hat{S}_{1_Z}|\Phi_n\ra
=\frac{1}{2}\sum_{n=1}^N\alpha_n,
\eeqa
where $\alpha_n=p_n^{(e)}-p_n^{(g)}$ is the probability difference 
for the excited and ground state in state $n$.

In Eq. (\ref{sec}), 
note that ${S}^{2}_{1_Z}={\rm\bf1}_{N}/4$ and therefore $\la\hat{S}^{2}_{1_Z}\ra=1/4$ for 
the normalized state $\Psi(x_{1},\cdots,x_{N})$.
The cross term can be evaluated as
\beqa
&&N(N-1)\la\hat{S}_{1_Z}\hat{S}_{2_Z}\ra
\nonumber\\
&=&\!\sum_{n,m}\!\left(\!\la\Phi_n|\hat{S}_{1_Z}|\Phi_n\ra\!\la\Phi_m|\hat{S}_{2_Z}|\Phi_m\ra
-\la\Phi_n\Phi_m|\hat{S}_{1_Z}\hat{S}_{2_Z}|\Phi_m\Phi_n\ra\!\right)\nonumber\\
&=&\frac{1}{4}\sum_{n,m}\left(\alpha_n \alpha_m-\Delta_{nm}\right),
\eeqa
where the $\Delta_{nm}$ terms are positive and defined as
\beqa
\Delta_{nm}&=&\left|\int {\rm d}x
[\phi_n^{(e)}(x)]^{*}\phi_m^{(e)}(x)-[\phi_n^{(g)}(x)]^{*}\phi_m^{(g)}(x)\right|^2
\nonumber\\
&=&\left|\la\phi_n^{(e)}|\phi_m^{(e)}\ra-\la\phi_n^{(g)}|\phi_m^{(g)}\ra \right|^2.
\eeqa
Combining these results, the variance simply reads
\beqa
(\Delta S_Z)^2=\frac{N}{4}-\frac{1}{4}\sum_{n,m}\Delta_{nm}.
\eeqa
If the dependence of single particle expectation value 
$\la\Phi_n|\hat{S}_{1_Z}|\Phi_n\ra$ on $n$ can be neglected, 
so that $\alpha_n\simeq \alpha$ for all $n$, 
\beqa
(\Delta S_Z)^2&=&\frac{N}{4}(1-\frac{1}{N}\sum_{n}\alpha_n^2)-\frac{1}{4}
\sum_{n,m\,\,n\neq m}\Delta_{nm}
\label{deltasz}\\
&\simeq&\frac{N}{4}(1-\alpha^2)-\frac{1}{4}\sum_{n,m\,\,n\neq m}\Delta_{nm}
\label{deltasz2}\\
&<& (\Delta S_Z)_0^2,
\eeqa
where we have identified a term $(\Delta S_Z)_0^2\equiv\frac{N}{4}(1-\alpha^2)$  
corresponding to 
the quantum noise for the Hartree product state in \cite{Itano93},  
and a negative correction for the strongly interacting Bosonic TG gas.  
\section{The Ramsey interferometer}
\begin{figure}
\begin{center} 
\includegraphics[angle=0,width=\linewidth]{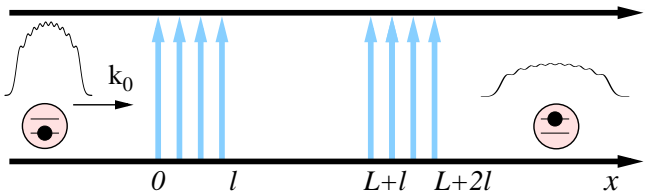}
\vspace*{.2cm}
\caption{\label{setup} Schematic setup for Ramsey interferometry 
of guided atoms in the spatial domain.
The atoms are prepared in the ground state and the probability of excitation 
is measured after passing the two fields.}
\end{center}
\end{figure}

Ramsey interferometry with guided ultracold atoms has recently been discussed in \cite{SeiMu-EPJD}.
Here we consider a system of $N$ two-level atoms in the Tonks-Girardeau regime,
initially confined in their ground internal states in a harmonic trap of frequency $\om$.
All energy scales are supposed to be much smaller than the transverse excitation energy $\hbar\om_{\perp}$, so that the radial degrees of freedom  are frozen out and the system is effectively one-dimensional.
The cloud is prepared in the ground state, and released by switching off the trap at time $t=0$;  
a momentum kick $\hbar k_0$ is also applied, so that the cloud moves along the $x$ axis 
towards the two separated oscillating fields localized between $0$ and $l$ and between $l+L$ and $2l+L$ (Fig.~\ref{setup}).
The initial state is prepared far from the first field. We thus have to take into account
the spatial width (root of the variance) of the highest state, $\delta_N=[(N+1/2)\hbar/(m\omega)]^{1/2}$, 
and choose the central initial position of the harmonic trap $x_0<0$ so that $x_0<<\delta_N$.      
   
In an oscillating-field-adapted interaction picture and using the Lamb-Dicke (see the next section), dipole and rotating-wave approximations 
the Hamiltonian is, for each of the particles,  
\begin{equation}
H = \frac{\widehat{p}^{\,2}}{2m} - 
\hbar\Delta \ket{e}\bra{e} + \frac{\hbar}{2}\Omega(\widehat{x}) (\ket{g}\bra{e} + \ket{e}\bra{g}),
\end{equation}
where the first term counts for the kinetic energy of the atom, $\Delta = \omega_L - \omega_{12}$ is the detuning between the oscillating field frequency and atomic transition frequency, and $\Omega(x)$ is the position-dependent Rabi frequency. For the explicit $x$ dependence we assume mesa functions, $\Omega(x) = \Omega$ for $x\in [0,l]$ and $x \in [l+L,2l+L]$ and zero elsewhere. In addition, we have to include the interparticle interactions but this is done implicitly by means of the wave function (\ref{bos})
and its boundary conditions at contact.    
%
%

%

%
The Ramsey pattern is defined by the dependence on the detuning 
of the probability of excited atoms after the interaction with the two field regions. 
From Eq. (\ref{sumn}) it follows that  $P_{e}^{(N)}=\sum_{n=1}^{N}P_{e}^{(n)}$, which is a 
remarkably simple result for an $N$-body system with external and interparticle interactions.  
Once a particle incident from the left and prepared in the state 
$e^{ik_0(x-x_0)}\phi_{n}(x-x_0)|g\ra$ at $t=0$ 
has passed completely through both fields, the probability amplitude for it to be in the 
excited state is 
\beqa
\phi_n^{(e)}(x,t)=\frac{1}{\sqrt{2\pi}}\int\d k e^{iqx-ik^2\hbar t/(2m)}T_{ge}(k)\tilde{\phi}_n(k),
\eeqa
where $\tilde{\phi}_n(k)$ is the wavenumber representation of the kicked $n$-th harmonic 
eigenstate, 
\beqa
\tilde{\phi}_n(k)&=&\frac{(-i)^n}{\sqrt{2^n n!}}\left(\frac{2\delta_0^2}{\pi}\right)^{1/4}
\nonumber\\
&\times&
e^{-\delta_0^2(k-k_0)^2}e^{-ik x_0}H_n[\sqrt{2}\delta_0(k-k_0)],
\eeqa
the momentum in the excited state is $q = \sqrt{k^2+2m\Delta/\hbar}$, the spatial width of the $n=0$ state is $\delta_0=[\hbar/(2 m\om)]^{1/2}$, $H_n$ the Hermite polynomials, and $T_{ge}$ is the
``double-barrier'' transmission amplitude for the 
excited state corresponding to atoms incident in the ground state (the excited state probability 
for monochromatic incidence in the ground state is 
$\frac{q}{k}|T_{ge}|^2$). 
The full quantum treatment of $T_{ge}$ can be done by means of the 
two-channel recurrence relations connecting it with one-field transmission 
and reflection amplitudes \cite{SeiMu-EPJD}.

\begin{figure}
\begin{center} 
\includegraphics[angle=0,width=\linewidth]{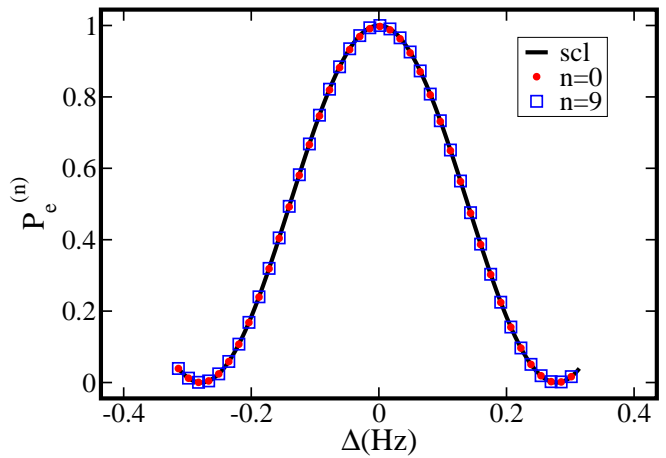}
\vspace*{.2cm}
\caption{\label{fringe} Central fringe for Ramsey interferometry in the spatial domain. 
The agreement is shown for the $n=0,9$ single-particle wavepackets and 
the semiclassical result, for $^{133}$Cs atoms, 
with $\hbar k_0/m=1$ cm/s, $l=1$ cm, $L=10$ cm, and $t=15$ s.}
\end{center}
\end{figure}


\begin{figure}
\begin{center} 
\includegraphics[angle=0,width=\linewidth]{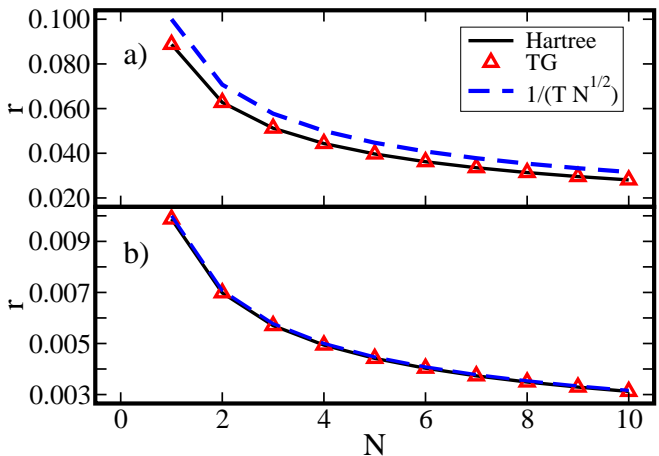}
\vspace*{.2cm}
\caption{\label{ratio} Quantum projection noise ratio, Eq. (\ref{r}), for the Ramsey 
interferometry in the spatial domain. The two-level Tonks-Girardeau gas ratio 
is essentially that of a Hartree product (uncorrelated atoms). 
The same parameters as in Fig. \ref{fringe} are used, with a) $L=10$ cm, b) $L=100$ cm.}
\end{center}
\end{figure}

Our numerical simulations are for $l=1$ cm, $L=10$ cm, $N=10$, and $v_0=1$ cm/s. 
Figure \ref{fringe} shows that the variation of the excitation probability for
different harmonic eigenstates is negligible in the scale shown, and in fact 
the curves for the central fringe are indistinguishable from the semiclassical 
result of Ramsey (which assumes classical motion for the center of mass,
uncoupled from the internal levels), 
\begin{eqnarray}
\label{eq:Ramsey_limit}
P_{12}(\Delta) &=& \frac{4\Omega^2}{\Omega'^2} \sin^2\left(\frac{\Omega' \tau}{2} \right) 
\bigg[\cos\left(\frac{\Omega' \tau}{2} \right) \cos\left(\frac{\Delta T}{2}  \right) 
\nonumber\\
&-& \frac{\Delta}{\Omega'} \sin\left(\frac{\Omega' \tau}{2} \right) \sin\left(\frac{\Delta T }{2}  \right) 
\bigg]^2,
\end{eqnarray}
where $\tau=l/v_0$, $T=L/v_0$, and $\Omega'=(\Omega^2+\Delta^2)^{1/2}$. 

There is 
in principle a broadening of the central fringe by increasing $n$ due to the
momentum broadening of vibrationally excited states. 
This effect may be expected however to be quite small for the few-body states
of our calculations, $N=10$, which is in fact of the order of
current experiments with TG gases ($N\approx 15, 50$ in \cite{Par04,Kin04})  
The width (root of the variance)
of the velocity distribution 
around the central velocity $v_0=\hbar k_0/m$ for the $n$-th state is
\beq
\Delta_v=\left[\left(n+\frac{1}{2}\right)\frac{\omega \hbar}{m}\right]^{1/2}=\sqrt{2n+1}\frac{\hbar}{{2}
m\delta_0},
\eeq
where we have used the spatial width of the $n=0$ state, $\delta_0=[\hbar/(2m\omega)]^{1/2}$.   
This will not affect significantly the width of the central fringe (proportional 
to the inverse of the crossing time $T$) as long as $\Delta_v/v_0<<1$.    
For $v_0=1$ cm/s, $\delta_0=20 \mu$m, the mass of $^{133}$Cs, and $N=10$, this ratio is $\sim 5\times 10^{-3}$. 
$N$ should be $\sim 4\times10^5$ to get a ratio of order one, but this means four orders of magnitude more particles than in the existing experiments.  
 
The error to estimate the atomic frequency from the Ramsey pattern depends on the ratio 
\beq
r=\frac{\Delta S_Z}{|\partial \la S_Z\ra/\partial\Delta|}
\label{r}
\eeq
which we calculate at half height of the central interference peak. 
We compute $\Delta S_Z$ with Eq.(\ref{deltasz}). 
Since, according to the previous discussion, the excitation probabilities are essentially independent of $n$, Eq.(\ref{deltasz2}) is an excellent approximation. 
Moreover, the correction to $(\Delta S_Z)_0$ due the particle correlations is negligible, with a relative error $[(\Delta S_Z)_0-\Delta S_Z]/\Delta S_Z\sim 10^{-10}$ in our calculations.      
Since, in addition, the derivative in Eq. (\ref{r}) is very well approximated by the semiclassical result, the ratio $r$ essentially coincides with that for freely moving, uncorrelated particles \cite{Itano93} and, for $L>>l$  it gives $1/(T\sqrt{N})$ for all $\Delta$, see Fig. \ref{ratio}.

\section{Ramsey interferometry in the time domain for guided atoms}
\begin{figure}[t]
\begin{center}
\includegraphics[height=3cm]{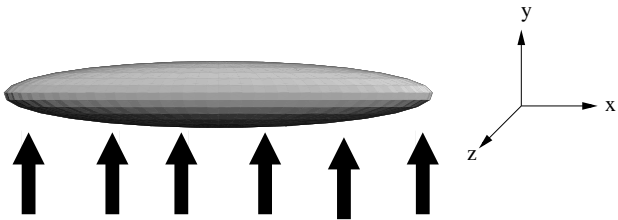}
\caption[]{Schematic setup for Ramsey interferometry in time domain. 
The TG gas is confined in a cigar-shaped trap and illuminated by a laser in $y$-direction.}
\label{cigar_trap}
\end{center}
\end{figure}

An alternative to the previous set-up is the separation of the fields 
in time rather than space but, at variance with the usual procedure, 
keeping the gas confined transversally at all times as required for the $1D$ regime of the 
TG gas, Fig. \ref{cigar_trap}. Because of the tight confinement the transverse vibrational excitation is 
negligible so that the Ramsey pattern is given by the standard expression
irrespective of the value of $n$. The whole TG gas therefore
produces the usual Ramsey pattern (\ref{eq:Ramsey_limit}) as we shall see in more detail.       
%
%

A two level atom in a cigar shape trap with characteristic frequencies $\omega_x,\omega_y$ and $\omega_z$ 
($\omega_x\ll\omega_y\sim\omega_z$) interacting with a (classical) laser field
directed in the perpendicular $y$ direction
is described (in a laser adapted interaction picture) by the Hamiltonian 
\begin{eqnarray}
\label{hamiltonian}
H&=&\sum_{i=x,y,z}\hbar\omega_i\left(a_i^\dag a_i+\frac{1}{2}\right)-{\hbar\Delta}|e\ra\la e|
\nonumber\\
& &+\frac{\hbar\Omega}{2}
\left[e^{i\eta_y\left(a_y+a_y^\dag\right)}\sigma_++H.c\right]
\end{eqnarray}
where $\sigma_+=|e\ra\la g|$. 
The Rabi frequency $\Omega$ is here a constant, independent of $x$, and $a_i^\dag$ ($a_i$) 
are the creation (annihilation) operators of the vibrational modes in the direction of the subscript. 
The parameter $\eta_y=k_L y_0$ is known as the Lamb-Dicke (LD) parameter, with 
$y_0=\sqrt{\hbar/2m\omega_y}$ being the extension of the atomic ground state in $y$-direction. 
The vibrational modes in the longitudinal $x$-direction are not coupled by the field
if the $x$-dependence of the field is negligible in the scale $\delta_N$ of the cloud. 
Also, motion in the $z$-direction remains uncoupled.

\subsection{Lamb-Dicke regime}

A particular interesting limit when dealing with trapped atoms interacting with laser fields 
is the so called Lamb-Dicke regime. In this regime, the extension of the atom's wave function in 
the direction of the field is much smaller than the laser wavelength, i.e., $\eta_y\ll1$.
If the LD regime is assumed, it is natural to approximate the exponentials in the coupling term
of the Hamiltonian (\ref{hamiltonian}) by $e^{\pm i\eta_y\left(a_y+a_y^\dag\right)}\approx1$, giving an 
approximate Hamiltonian
\begin{eqnarray}
\label{hamiltonian_LD}
H_{LD}&=&\sum_{i=x,y,z}\hbar\omega_i\left(a_i^\dag a_i+\frac{1}{2}\right)
-\hbar\Delta|e\ra\la e|\nonumber\\
& &+\frac{\hbar\Omega}{2}
\left(\sigma_++\sigma_-\right)
\end{eqnarray}
which does not couple the vibrational modes in the transversal $y$-direction. 
In this regime, the vibrational levels are well separated and the fields cannot induce transitions between them
(the recoil frequency is much smaller than the trapping frequency). 
Within this approximation the number operators $n_i=a_i^\dag a_i$ are some constant of motion for $i=x,y,z$ and thus the dynamics of the system is independent of the vibrational modes, 
reproducing the usual Ramsey fringe pattern (\ref{eq:Ramsey_limit}) when time separated pulses are applied.

%
%
%

\subsection{TG regime}

A tight transversal confinement is needed in order to reach the TG regime, which
is achieved when the dimensionless parameter $\gamma=mg/\hbar^2n\gg1$. Here 
$n\approx N/\delta_N$ is the linear density of the gas  
and $g$ the 1D interaction strength, which is  
given by $g=2\hbar^2 a/m y_0^2$, with $a$ being the s-wave scattering length \cite{RT04}. 
We may then write the criterion for being in the TG regime as $2a/ny_0^2\gg1$ or 
\beqa
\label{TG_condition}
\gamma=\frac{2a\delta_N}{Ny_0^2}\gg1.
\eeqa

If $^{133}$Cs atoms in a trap with frequency
$\omega_y\approx2\pi\times 1$MHz are considered, the transversal confinement turns out to be $y_0\approx 6$nm. For 
$\delta_0\approx20\mu$m,  a scattering length of $a\approx100a_0$, $a_0$ being the Bohr radius, and $N=10$,  $\gamma\approx2\times10^3$, well in the TG regime.

If the hyperfine transition of the $^{133}$Cs atom
at $9.192$GHz is driven, a LD parameter of $\eta_y=k_L y_0\approx10^{-6}$ is obtained for the transversal confinement of
$y_0\approx6$nm previously estimated, which is well inside the LD regime. The TG condition then imposes the 
LD condition for microwave transitions. For optical transitions, LD parameters of $0.05-0.1$ are obtained, 
that can also be considered to lay into the LD regime.
\\

\section{Summary and discussion}

A model of $N$ Bosons in 1D with contact interactions that interchange the 
momentum and internal state of the 2-level atoms subjected to two oscillating fields 
has been worked out. 

A word is in order about the interactions between different channels. In the singlet component, 
the space wavefunction is antisymmetric, so that s-wave scattering is forbidden, 
and the interactions are governed to leading order by a 3D p-wave scattering amplitude. 
In close analogy with spin-polarized Fermions, such 
interactions can be enhanced by a p-wave Feschbach resonance, 
with an associated 1D odd-wave confinement-induced one-dimensional 
Feschbach resonance (CIR) which allows to engineer $v_s$ \cite{GirOls03,GirOls04,GirNguOls04,GB04}.

For realistic parameters in the ultracold regime the 
system behaves similarly for spatial or temporal separation of the fields, and 
according to the semiclassical Ramsey pattern for independent, freely moving 
particles. 
Moreover, the quantum projection noise reminds close to that of an ensemble of independent atoms.

For the two-level Tonks-Girardeau gas, the interactions do not worsen 
the quality of the Ramsey pattern but have the additional advantage 
of dramatically reducing the three-body correlation function \cite{GanShl03,DMG07} and 
therefore enhancing the stability of the gas with respect to the ideal case.
We expect strongly interacting gases to play a remarkable role in 
interferometry with ultracold atoms in waveguides.

\section*{Acknowledgements}
We acknowledge discussions with M. Girardeau.
This work has been supported by Ministerio de Edu\-ca\-ci\'on y Ciencia (FIS2006-10268-C03-01) and UPV-EHU (00039.310-15968/2004).
S. V. M. acknowledges a research visitor Ph. D. student fellowship  
by the Ministry of Science, Research and Technology of Iran.
A. C. acknowledges financial support by the Basque Government (BFI04.479).

\end{document}